\documentclass[seceq]{ptptex}
\newcommand{\bfg}{ \boldsymbol{g} }

\newcommand{\bfA}{ \boldsymbol{A} }
\newcommand{\bfB}{ \boldsymbol{B} }
\newcommand{\bfC}{ \boldsymbol{C} }
\newcommand{\bfD}{ \boldsymbol{D} }

\newcommand{\bfG}{ \boldsymbol{G} }

\newcommand{\bfK}{ \boldsymbol{K} }
\newcommand{\bfL}{ \boldsymbol{L} }
\newcommand{\bfM}{ \boldsymbol{M} }

\newcommand{\bfS}{ \boldsymbol{S} }

\newcommand{\bfeps}{\boldsymbol{\epsilon}}
\newcommand{\bfsig}{\boldsymbol{\sigma}}
\newcommand{\vecp}{ {\vec p} }

\newcommand{\vecx}{ {\vec x} }

\newcommand{\vecA}{ {\vec A} }
\newcommand{\vecB}{ {\vec B} }
\newcommand{\vecC}{ {\vec C} }

\newcommand{\vecE}{ {\vec E} }

\newcommand{\vecJ}{ {\vec J} }

\newcommand{\vecP}{ {\vec P} }

\newcommand{\vecX}{ {\vec X} }

\newcommand{\veclam}{ {\vec \lambda} }

\newcommand{\tilE}{ \tilde{E} }
\newcommand{\tilF}{ \tilde{F} }

\newcommand{\vectilE}{ \vec{\tilE} }
\newcommand{\vectilF}{ \vec{\tilF} }
\newcommand{\tilbfM}{ \tilde{\bfM} }

\newcommand{\sgn}{{\mbox{sgn}}}
\newcommand{\Tr}{\mbox{Tr}}

\newcommand{\DEF}{:=}
\notypesetlogo                       
\title{%        %You can use \\ for explicit line-break
Noncommutative Hall Effect
}

\author{%       %Use \scshape  for the family name
    Akira \textsc{Kokado}$^{1,}$\footnote{
                         E-mail: kokado@kobe-kiu.ac.jp},
    Takashi \textsc{Okamura}$^{2,}$\footnote{
                         E-mail: okamura@ksc.kwansei.ac.jp}
and Takesi \textsc{Saito}$^{3,}$\footnote{
                         E-mail: tsaito@k7.dion.ne.jp}
}
\inst{%         %Affiliation, neglected when [addenda] or [errata]
$^1$Kobe International University, Kobe 658-0032, Japan
\\
$^{2,3}$Department of Physics, Kwansei Gakuin University,
    Sanda 669-1337, Japan
}

\abst{%         %this abstract is neglected when [addenda] or [errata]
When coordinates are noncommutative, the Hall effect is reinvestigated.
The Hall conductivity is expressed with noncommutative parameters,
so that in the commutative limit it tends to the conventional result.
}
\begin{document}
\maketitle
%%%%%%%%%%%%%%%%%%%%%%% Section 1 %%%%%%%%%%%%%%%%%%%%%%%%%%%
\section{Introduction}
Recently the idea of noncommutative structure at small length scales
has been drawn much attention in string theories and
field theories including gravity.
The general expectation was that noncommutative spacetimes could
introduce an effective cut-off in field theories,
in analogy to a lattice. 
Especially interesting is a model of open strings propagating
in a constant $B$ field background.
Previous studies show that this model is related to
noncommutativity of D-branes,
\cite{rf:1} \ 
and in the zero slope limit
to noncommutative Yang-Mills theory.
\cite{rf:2} \ 
An intriguing mixing of UV and IR theories has been also found
in the perturbative dynamics of noncommutative field theories.
\cite{rf:3}

In this paper we consider the Hall effect in noncommutative spaces,
in order to clarify what a role of noncommutative coordinates plays
in this phenomenon.
This model would provide the most simple example of
noncommutative quantum mechanics.
We consider here a system that a particle with charge (-$e$ ) moves
on a two-dimensional plane in the uniform external electric field
$\vecE$ and the uniform external magnetic field $\vecB$,
where the direction of $\vecB$ is parallel to the $(x^1, x^2)$ plane,
while that of $\vecB$ is transverse to the $(x^1, x^2)$ plane.
Let coordinates of the particle be $\vecx = (x^1, x^2)$ and
their conjugate momenta be $\vecp = (p_1, p_2)$.
This system is described by the Hamiltonian
\begin{equation}
\label{101}
 H = \frac{1}{2m}( \vecp +e \vecA )^2 - e\phi~.
\end{equation}
In the following we choose the gauge
\begin{equation}
\label{102}
 e \vecA = \bfA~\vecx~,
\end{equation}
where
\begin{eqnarray}
 & & \bfA = \begin{pmatrix} 
                   a_{11} & a_{12} \\
                   a_{21} & a_{22}
            \end{pmatrix}~,
\quad 
 eB_3 = a_{21}-a_{12}~,
\label{103}
\end{eqnarray}
and
\begin{equation}
\label{104}
 \phi = -\vecE \cdot \vecx~.
\end{equation}
Hence, the Hamiltonian is written as
\begin{equation}
\label{105}
 H = \frac{1}{2m}\left(p_i+a_{ij}x^j\right)^2 + eE_{i}x^i.
\end{equation}
\indent
By noncommutative spaces we mean that we have the commutation relation
(CR) for coordinates, $\left[ x, y\right]=i\hbar \theta $,
where $\theta $ is a real parameter.
Generally one can set for our system a CR in matrix form
\cite{rf:4}
\begin{eqnarray}
 & & \omega \DEF
    \begin{pmatrix}
         [x^1,x^1] & [x^1,x^2] & [x^1,p_1] & [x^1,p_2] \\
         [x^2,x^1] & [x^2,x^2] & [x^2,p_1] & [x^2,p_2] \\
         [p_1,x^1] & [p_1,x^2] & [p_1,p_1] & [p_1,p_2] \\
         [p_2,x^1] & [p_2,x^2] & [p_2,p_1] & [p_2,p_2]
    \end{pmatrix}
   = i \hbar~\begin{pmatrix}
             \theta_x~\bfeps & \bfg^T \\
                      - \bfg & \theta_p~\bfeps \end{pmatrix}~,
\label{106} \\
 & & \bfeps := \begin{pmatrix}  0 & 1 \\ -1 & 0 \end{pmatrix}~,
\nonumber
\end{eqnarray}
parametrized by real numbers $\theta _x$,$\theta _y$ 
and real $2\times 2$  matrix $\bfg $.
The Hamiltonian (\ref{105}) is understood to be symmetrized
with respect to variables. \\
\indent
In the following sections we calculate the Hall conductivity
by using the Hamiltonian (\ref{105}) and
the noncommutative CR (\ref{106}).
In Sec.2 a case of $\det \omega \neq 0$ is considered,
while the other case  $\det \omega = 0$ is considered in Sec.3.
The final section is devoted to concluding remarks.
The noncommutative Hall effect was also considered
quantum-mechanically
in Ref.~\citen{rf:5}, \ while field-theoretically
in Ref.~\citen{rf:6}. \ However, they were interested especially
in noncommutativity of coordinates.
Our consideration, on the other hand,
includes more general noncommutative CR's of the type Eq.(\ref{106}).
%%%%%%%%%%%%%%%%%%%%%%% Section 2 %%%%%%%%%%%%%%%%%%%%%%%%%%%
\section{A case of $\det \omega \neq 0$}
In this section we consider the case of $\det \omega \neq 0$.
By using the formula for block matrices 
\begin{equation}
 \det \begin{pmatrix} \bfA & \bfB \\ \bfC & \bfD \end{pmatrix}
= \left( \det\bfA \right)~\det\left( \bfD - \bfC\bfA^{-1}\bfB \right)
= \left( \det\bfD \right)~\det\left( \bfA - \bfB\bfD^{-1}\bfC \right)~,
\label{201}
\end{equation}
one obtains
\begin{equation}
 \det \omega = (i \hbar)^4
          \left( \theta_x \theta_p - \det \bfg \right)^2~ \neq 0.
\label{202}
\end{equation}
Let us write the Hamiltonian (\ref{105}) in the form 
\begin{eqnarray}
 H = \frac{1}{2m} \begin{pmatrix} \vecx^T & \vecp^T \end{pmatrix} \bfM
                  \begin{pmatrix} \vecx \\ \vecp \end{pmatrix}
      + e \begin{pmatrix} \vecx^T & \vecp^T \end{pmatrix}
                \begin{pmatrix} \vecE \\ \vec{0} \end{pmatrix}~,
\label{203}
\end{eqnarray}
where 
\begin{eqnarray}
  \vecx = \begin{pmatrix} x^1 \\ x^2 \end{pmatrix}~, \hspace{1cm}
  \vecp = \begin{pmatrix} p_1 \\ p_2 \end{pmatrix}~,
\label{204}
\end{eqnarray}
\begin{eqnarray}
 & & \bfM = \begin{pmatrix} \bfA & \boldsymbol{1} \\
                    \boldsymbol{0} & \boldsymbol{0} \end{pmatrix}^T~
             \begin{pmatrix} \bfA & \boldsymbol{1} \\
                    \boldsymbol{0} & \boldsymbol{0} \end{pmatrix}
          = \begin{pmatrix} \bfA^T~\bfA & \bfA^T \\
                             \bfA & \boldsymbol{1} \end{pmatrix}~,
\label{205}
\end{eqnarray}
$A$ being given by Eq.(\ref{103}).
The Heisenberg equations of motion for $\vec{x}$ and $\vec{p}$
are given by
\begin{eqnarray}
   \frac{d}{dt} \begin{pmatrix} \vecx \\ \vecp \end{pmatrix}
  = \frac{1}{i \hbar}
      \begin{pmatrix} \left[~\vecx,~H~\right] \\
                      \left[~\vecp,~H~\right] \end{pmatrix}~.
\label{206}
\end{eqnarray}

\indent
In order to solve this equation (\ref{206}),
we consider a linear transformation of variables
\begin{equation}
  \begin{pmatrix} \vecx \\ \vecp \end{pmatrix}
    =  S \begin{pmatrix} \vecX \\ \vecP \end{pmatrix}
         + \begin{pmatrix} \veclam_x \\ \veclam_p \end{pmatrix}~,
\label{207}
\end{equation}
where
\begin{eqnarray}
 S &\DEF& \begin{pmatrix} \bfS_{xx} & \bfS_{xp} \\
                          \bfS_{px} & \bfS_{pp} \end{pmatrix}
 = \begin{pmatrix} \alpha \bfA & \alpha \boldsymbol{1} \\
                    \beta \bfB & \beta \bfD \end{pmatrix}^{-1}
 \nonumber \\
  &=& \begin{pmatrix}
       -(\beta/\alpha) \bfS_{xp} \bfD
                 & (1/\beta) \left( \bfB - \bfD \bfA \right)^{-1} \\
       \left( \boldsymbol{1} + \beta \bfA \bfS_{xp} \bfD \right)/\alpha
                 & - \bfA \bfS_{xp} \end{pmatrix}
\label{208}
\end{eqnarray}
and its determinant is assumed not to be zero, i.e.,
\begin{eqnarray}
 \det \begin{pmatrix} \alpha \bfA & \alpha \boldsymbol{1} \\
                       \beta \bfB & \beta \bfD \end{pmatrix}
    = ( \alpha \beta )^2 \det\left( \bfB - \bfD \bfA \right) \neq 0~.
\label{209}
\end{eqnarray}
Here, $2\times 2$  matrices $\bfB$, $\bfD$ and
parameters $\alpha $, $\beta $, $\veclam$~'s will be fixed later.
In terms of the new variables $\vec{X}$, $\vec{P}$,
the Hamiltonian (\ref{203}) can be written as
\begin{eqnarray}
  H &=& \frac{1}{2m} \begin{pmatrix} \vecX^T & \vecP^T \end{pmatrix}
         \tilbfM     \begin{pmatrix} \vecX \\ \vecP \end{pmatrix}
       + e \begin{pmatrix} \vecX^T & \vecP^T \end{pmatrix}
           \begin{pmatrix} \vectilE \\ \vectilF \end{pmatrix}
\nonumber \\
 & & + \frac{1}{2m} \left\vert~\veclam_p + \bfA \veclam_x~\right\vert^2
 + e \vecE \cdot \veclam_x~, 
\label{210}
\end{eqnarray}
where
\begin{eqnarray}
 \tilbfM &\DEF&
   \left[ \begin{pmatrix} \bfA & \boldsymbol{1} \\
                \boldsymbol{0} & \boldsymbol{0} \end{pmatrix}
 \begin{pmatrix} \alpha \bfA & \alpha \boldsymbol{1} \\
                  \beta \bfB & \beta \bfD \end{pmatrix}^{-1} \right]^T
   \left[ \begin{pmatrix} \bfA & \boldsymbol{1} \\
                \boldsymbol{0} & \boldsymbol{0} \end{pmatrix}
 \begin{pmatrix} \alpha \bfA & \alpha \boldsymbol{1} \\
                  \beta \bfB & \beta \bfD \end{pmatrix}^{-1} \right]
\label{eq:def-tilbfM} \nonumber  \\
  &=& \begin{pmatrix} \boldsymbol{1}/\alpha & \boldsymbol{0} \\
                             \boldsymbol{0} & \boldsymbol{0}
                                                       \end{pmatrix}^T
      \begin{pmatrix} \boldsymbol{1}/\alpha & \boldsymbol{0} \\
                             \boldsymbol{0} & \boldsymbol{0}
                                                       \end{pmatrix}
   =  \begin{pmatrix} \boldsymbol{1}/\alpha^2 & \boldsymbol{0} \\
                               \boldsymbol{0} & \boldsymbol{0}
                                                       \end{pmatrix}~,
\label{211}
\end{eqnarray}
and
\begin{eqnarray}
 \begin{pmatrix} \vectilE \\ \vectilF \end{pmatrix} &\DEF&
      \begin{pmatrix} \bfS_{xx} & \bfS_{xp} \\
                      \bfS_{px} & \bfS_{pp} \end{pmatrix}^T
 \left[~ \begin{pmatrix} \vecE \\ \vec{0} \end{pmatrix}
    + \frac{1}{m~e} \begin{pmatrix}
                          \bfA^T \bfA & \bfA^T \\
                                 \bfA & \boldsymbol{1} \end{pmatrix}
      \begin{pmatrix} \veclam_x \\ \veclam_p \end{pmatrix}~\right]~,
\label{eq:def-tilEF} \nonumber \\
  &=& \begin{pmatrix} \bfS_{xx}^T \vecE \\
                      \bfS_{xp}^T \vecE \end{pmatrix}
    + \frac{1}{m e~\alpha}
        \begin{pmatrix} \bfA \veclam_x + \veclam_p \\
                                          {\vec 0} \end{pmatrix}~.
\label{212}
\end{eqnarray}
From Eq.(\ref{212}) one can see that the transformed electric field
$\vectilE$ is always made to be zero if $\veclam_p$ is properly chosen.
In this case the Hamiltonian (\ref{210}) is reduced to
\begin{eqnarray}
  H = \frac{1}{2m \alpha^2}  \vecX^2
    + e \vecP^T~\left( \bfS_{xp}^T~\vecE \right)~,
\label{213}
\end{eqnarray}
except for constant factors.

On the other hand, the CR in matrix form $\omega$ for old variables
is transformed into $\Omega$ for new variables $\vecX$ and $\vecP$
under the transformation Eq.(\ref{207}):
\begin{eqnarray}
 \Omega &=&
     \begin{pmatrix}
        \alpha \bfA & \alpha \boldsymbol{1} \\
         \beta \bfB & \beta \bfD     \end{pmatrix} \omega~
     \begin{pmatrix}
        \alpha \bfA & \alpha \boldsymbol{1} \\
         \beta \bfB & \beta \bfD \end{pmatrix}^T
  = i \hbar~\begin{pmatrix}
               \theta_X \bfeps & \bfG^T \\
                        - \bfG & \theta_P \bfeps \end{pmatrix}~,
\label{214}
\end{eqnarray}
where
\begin{eqnarray}
 & & \bfG = \alpha \beta~\left[
       \bfD \left( \bfg \bfA^T - \theta_p \bfeps \right)
     - \bfB \left( \bfg - \theta_x \bfA \bfeps \right)^T \right]~,
\label{215} \\
 & & \theta_X = \alpha^2~\zeta~, \hspace{1cm} \theta_P = \beta^2~\eta~,
\label{216}
\end{eqnarray}
with
\begin{eqnarray}
 & & \zeta = \theta_p + (\det\bfA) \theta_x
      + \Tr\left( \bfg \bfA^T \bfeps \right)~,
\label{217} \\
 & & \eta =  (\det\bfD) \theta_p + (\det\bfB) \theta_x
      + \Tr\left( \bfD \bfg \bfB^T \bfeps \right)~.
\label{218}
\end{eqnarray}
Here we have used the formula $\displaystyle{
\bfeps~\bfL~\bfeps~ \bfL^T = - ( \mbox{det} \bfL ) \boldsymbol{1} }$
for an arbitrary $2\times 2$ matrix  $\bfL$.
Since $\omega /(i\hbar )$ is an antisymmetric real matrix,
it is generally transformed into the canonical form with
$\bfG= \boldsymbol{0}$ in Eq.(\ref{214}).
Actually, this can be seen in the following:
From Eq.(\ref{215}) one can see that it is always possible
to make $\bfG$ vanish, that is,
we can always put the following equality:
\begin{eqnarray}
   \bfD \left( \bfg \bfA^T - \theta_p \bfeps \right)
   =\bfB \left( \bfg - \theta_x \bfA \bfeps \right)^T~.
\label{219}
\end{eqnarray}
In fact, when $\bfK \DEF  \bfg - \theta_x \bfA \bfeps$ or
$\bfL \DEF  \bfg \bfA^T - \theta_p \bfeps$ is the regular matrix,
we always get $\bfD$ or $\bfB$  which satisfies Eq.(\ref{219}).
When $\bfK$ and $\bfL$ are both irregular matrices,
we obtain an equality $\theta_p = \theta_x \det\bfA$ or $\zeta=0$
from equations,
\begin{eqnarray}
 \det \bfK
  &=& \theta_x \zeta - \left( \theta_x \theta_p - \det \bfg \right)=0~,
\nonumber \\
 \det \bfL &=& \theta_p \zeta - \left( \det \bfA \right)
            \left( \theta_x \theta_p - \det \bfg \right)=0~.
\label{220}
\end{eqnarray}
Here we have made use of the formula, $\det\bfK = (1/2)\left[
\left( \Tr \bfK \right)^2 - \Tr\left( \bfK^2 \right)\right]$,
for an arbitrary $2\times 2$ matrix $\bfK$.
In the following we assume to exclude such a special cases
$\theta_p = \theta_x \det\bfA$ or $\zeta=0$.

In conclusion we see that it is generally possible to set
$\bfG = \boldsymbol{0}$ in $\Omega$.
Since $\det\Omega \ne 0$, it follows that $\theta_X,~\theta_P \ne 0$.
Hence, one can chooses $\alpha=|\zeta|^{-1/2}$,
$\beta = |\eta|^{-1/2}$ from Eqs.(\ref{216}).
Accordingly we have
\begin{eqnarray}
 \Omega &=& i \hbar
      \begin{pmatrix}
         \sgn(\zeta) \bfeps & \boldsymbol{0} \\
                    \boldsymbol{0} & \sgn(\eta) \bfeps \end{pmatrix}~,
\label{221}
\end{eqnarray}
to give
\begin{equation}
\label{222}
 \left[X^1, X^2 \right] = i\hbar~\sgn(\zeta)~, \quad
 \left[P_1, P_2 \right] = i\hbar~\sgn(\eta)~, \quad
 \left[X^i, P_j \right] = 0~.
\end{equation}
The Hamiltonian (\ref{213}) becomes
\begin{eqnarray}
  H = \frac{|\zeta|}{2m} \left\vert \vecX \right\vert^2
    + e \vecP^T~\left(\bfS_{xp}^T~\vecE \right)~.
\label{223}
\end{eqnarray}
In view of both equations (\ref{222}) and (\ref{223}),
we see that the first part in Eq.(\ref{223}) corresponds to
the harmonic oscillator part, and this is dynamically independent of
the second linear part.

The eigenvalue equation for the linear part is given by
\begin{equation}
 \left( C^1P_1 + C^2P_2 \right) \Psi _k = E_k \Psi _k~,
\label{224}
\end{equation}
where $\vecC = e~(S_{xp}^{\ \ T}\vec{E})$. \\
From Eq.(\ref{222}), one can use the representation
$P_2=-i\hbar~\sgn(\eta)~ \partial/\partial P_1$.
This is substituted into Eq.(\ref{224}) to find a solution
\begin{equation}
 \Psi _k = N_k \exp\left[\frac{\sgn(\eta)~C^1}{2i\hbar C^2}
     \left(P_1-\frac{k}{C^1}\right)^2\right]~,
\label{225}
\end{equation}
and then we obtain $E_k=k$, where $k$ is a real number.
Here, the eigenfunction $\Psi _k$ is normalized by the delta function
\begin{equation}
 \left(\Psi _k, \Psi _{k'}\right) = \delta \left( k - k' \right)~.
\label{226}
\end{equation}

Now, the Heisenberg equation (\ref{206}) is given,
in terms of new variables $\vec{X}$, $\vec{P}$, as
\begin{eqnarray}
   \frac{d}{dt} \begin{pmatrix} \vecx \\ \vecp \end{pmatrix}
 = S \frac{d}{dt} \begin{pmatrix} \vecX \\ \vecP \end{pmatrix}
 = S \frac{1}{i\hbar }
   \begin{pmatrix}
      \left[\vecX, H\right] \\ \left[\vecP, H\right] \end{pmatrix}
 = S~\frac{\Omega}{i \hbar}~
 \begin{pmatrix} |\zeta| \vecX/m \\ e \bfS_{xp}^T \vecE \end{pmatrix}~.
\label{227}
\end{eqnarray}
that is,
\begin{eqnarray}
 \frac{d \vecx}{dt} 
  &=& \frac{\zeta}{m} \bfS_{xx} \bfeps \vecX
     + \frac{e \eta}{ \det( \bfB - \bfD \bfA ) }~\bfeps \vecE~, 
\label{228} \\
 \frac{d \vecp}{dt} 
  &=& \frac{\zeta}{m} \bfS_{px} \bfeps \vecX
     - \frac{ e \eta }{ \det( \bfB - \bfD \bfA ) }~\bfA \bfeps \vecE~.
\label{229}
\end{eqnarray}
Let us consider the expectation value of Eq.(\ref{228})
for a stationary state of the Hamiltonian (\ref{223}),
which is a product state of the harmonic oscillator eigenfunction and
$\Psi _k$ given in Eq.(\ref{225}).
Since the expectation value of $\vecX$ is zero, i.e.,
$\langle \vecX \rangle = \vec{0}$, we have
\begin{equation}
  \left\langle \frac{d \vecx}{dt} \right\rangle
  = \frac{ e \eta }{ \det( \bfB - \bfD \bfA ) }~\bfeps \vecE~,
\label{230}
\end{equation}
Defining the electric current by
$\vecJ = (- e) \rho \langle {\dot \vecx} \rangle$ with $\rho$
the charge density, and the Hall conductivity
$\bfsig$ by $\vecJ  = \bfsig~\vecE$, we have
\begin{eqnarray}
 \bfsig = \frac{ - e^2 \rho \eta }{ \det( \bfB - \bfD \bfA ) }~\bfeps
 = \frac{ e^2 \rho \left( \det \bfg - \theta_x \theta_p \right) }
{\theta_p + (\det\bfA) \theta_x + \Tr\left( \bfg \bfA^T \bfeps \right)
   }~\bfeps~.
\label{231}
\end{eqnarray}
Here we have used the relation
\begin{eqnarray}
 \zeta~\eta &=& \left( \theta_x \theta_p - \det \bfg \right)
                 \det( \bfB - \bfD \bfA )~,
\label{232}
\end{eqnarray}
which comes from Eq.(\ref{219}).
%%%%%%%%%%%%%%%%%%%%%%% Section 3 %%%%%%%%%%%%%%%%%%%%%%%%%%%
\section{A case of $\det \omega = 0$}
In the case of $\det \omega = 0$ we have $\det \omega
= (i \hbar)^4 \left( \theta_x \theta_p - \det \bfg \right)^2=0$.
Since $\det \bfK = \theta_x \zeta$ and $\det \bfL = \theta_p \zeta$,
it is again possible generally to make $\bfG$ vanish,
provided that $\omega$ is not zero matrix.
Hence, it follows that $\det\Omega = (i \hbar)^4 \theta_X \theta_P =0$.
In the following we consider the case $\theta_X\neq 0$, $\theta_P=0$
\footnote{
In the case that $\theta_X =0$ and $\theta_P \ne 0$, we also
conclude that the Hall conductivity vanishes
along the similar argument for this case.
}.
In this case, $\vecP$ is commutable with $\vecX$ and $\vecP$,
hence, it should be c-number. The second term in the Hamiltonian
(\ref{223}) becomes a constant, so that the Hamiltonian describes
a single harmonic oscillator.
The equation of motion (\ref{227}) becomes
\begin{equation}
   \frac{d}{dt} \begin{pmatrix} \vecx \\ \vecp \end{pmatrix}
  = S \begin{pmatrix} \sgn(\zeta) \bfeps & \boldsymbol{0} \\
                          \boldsymbol{0} & \boldsymbol{0}
                                                      \end{pmatrix}~
   \begin{pmatrix} |\zeta| \vecX/m \\ \vec{0} \end{pmatrix}~.
\label{301}
\end{equation}
hence, 
\begin{equation}
  \frac{d \vecx}{dt} 
  = \frac{\zeta}{m} \bfS_{xx} \bfeps \vecX~.
\label{302}
\end{equation}
Its expectation value for the stationary state of the Hamiltonian
is zero. This means that we have no Hall current.

The physical meaning of this result is clear: Since
our system is constrained, the Hamiltonian in two-dimensional space
is reduced to the harmonic oscillator Hamiltonian
in one-dimensional space.
Hence the charged particle can move only around
the origin of the harmonic oscillator,
leading to zero-expectation values of currents.
%%%%%%%%%%%%%%%%%%%%%%% Section 4 %%%%%%%%%%%%%%%%%%%%%%%%%%%
\section{Concluding remarks}
We have considered the Hall effect when coordinates are noncommutative,
generally the commutation relations are given by Eq.(\ref{106}).
We have neglected the effect coming from particle spins,
since this effect is irrelevant to our purpose.
The Hall conductivity is given by Eq.(\ref{231}),
which depends on noncommutative parameters.
In the usual \lq\lq commutative" limit,
$\theta _x, \theta _p \to 0$, $g_{ij} \to \delta _{ij}$,
the Hall conductivity (\ref{231}) tends to the ordinary results,
\begin{equation}
 \sigma _{21}=\frac{e\rho }{B}, \quad \sigma _{11}=\sigma _{22}=0~.
\label{401}
\end{equation}
The formula (\ref{231}) will serve to constrain
the noncommutative parameters  $\theta _x$ and $\theta _p$
from experiments.
%%%%%%%%%%%%%%%%%%%%%% Acknowledgments %%%%%%%%%%%%%%%%%%%%%%
\section*{Acknowledgements}
We are grateful to Reijiro Kubo and Gaku Konisi
for their critical comments.
%
%\appendix
%\section{First Appendix} %Empty argument \section{} yields `Appendix'.
%
%\section{Second Appendix}
%%%%%%%%%%%%%%%%%%%%%%%%%%%%%%%%%%%%%%%%%%%%%%%%%%%%%%%%%%%%%
% Some macros are available for the bibliography:
%  o for general use
%    \JL : general journals                 \andvol : Vol (Year) Page
%  o for individual journal 
%    \AJ   : Astrophys. J.           \NC         : Nuovo Cim.
%    \ANN  : Ann. of Phys.           \NPA, \NPB  : Nucl. Phys. [A,B]
%    \CMP  : Commun. Math. Phys.     \PLA, \PLB  : Phys. Lett. [A,B]
%    \IJMP : Int. J. Mod. Phys.      \PRA - \PRE : Phys. Rev. [A-E]     
%    \JHEP : J. High Energy Phys.    \PRL        : Phys. Rev. Lett.
%    \JMP  : J. Math. Phys.          \PRP        : Phys. Rep.
%    \JP   : J. of Phys.             \PTP        : Prog. Theor. Phys.     
%    \JPSJ : J. Phys. Soc. Jpn.      \PTPS       : Prog. Theor. Phys. Suppl.
% Usage:
%  \PRD{45,1990,345}          ==> Phys.~Rev.\ \textbf{D45} (1990), 345
%  \JL{Nature,418,2002,123}   ==> Nature \textbf{418} (2002), 123
%  \andvol{B123,1995,1020}    ==> \textbf{B123} (1995), 1020
%%%%%%%%%%%%%%%%%%%%%%%%%%%%%%%%%%%%%%%%%%%%%%%%%%%%%%%%%%%%%
%%%%%%%%%%%%%%%%%%%%%%%% references %%%%%%%%%%%%%%%%%%%%%%%%%

%%%%%%%%%%%%%%%%%%%%%%%%%%%%%%%%%%%%%%%%%%%%%%%%%%%%%%%%%%%%%
\end{document}